\documentclass[12pt]{iopart}
\usepackage{iopams}  
\usepackage{graphicx, caption}
\usepackage{subcaption}
\usepackage{lipsum} 
\usepackage[normalem]{ulem}
\usepackage{xcolor}

\def\pT{p_{_T}}

\def\Raa{Q_\mathrm{AA}}
\def\qavg{\langle \hat q \rangle}

\newcommand{\beq}{\begin{eqnarray}}
\newcommand{\eeq}{\end{eqnarray}}
\newcommand{\be}{\begin{eqnarray*}}
\newcommand{\ee}{\end{eqnarray*}}
\newcommand{\bal}{\begin{align}}
\newcommand{\eal}{\end{align}}
\newcommand{\dd}{{\mathrm d}}

\begin{document}

\title{Jet quenching in expanding medium}
\author{S P Adhya , C A Salgado, M Spousta and K Tywoniuk}
\address{Institute of Particle and Nuclear Physics, Faculty of Mathematics and Physics, Charles University, V Holesovickach 2, 180 00 Prague 8, Czech Republic}
\address{Instituto Galego de F\'isica de Altas Enerx\'ias IGFAE, Universidade de Santiago de Compostela, E-15782 Galicia-Spain}
\address{Department of Physics and Technology, University of Bergen, 5007 Bergen, Norway}

\ead{souvik@mff.cuni.cz}
\vspace{10pt}
\begin{indented}
\item[]February, 2020
\end{indented}

\begin{abstract}
Comprehensive understanding of medium-induced radiative energy loss is of a paramount importance in describing observed jet quenching in heavy-ion collisions. In this work, we have calculated the medium-modified gluon splitting rates for different profiles of the expanding partonic medium, namely profiles for static, exponential, and Bjorken expanding medium. Here in this study, we have used the Baier-Dokshitzer-Mueller-Peigne-Schiff-Zakharov (BDMPSZ) formalism is used for multiple soft scatterings with a time-dependent transport coefficient for characterizing the expanding medium. Using the kinetic rate equation, we have numerically evaluated the distribution of the medium evolved gluon spectra for different medium profiles. We have presented a calculation of the jet $\Raa$ which quantifies a sensitivity of the inclusive jet suppression on the way how the medium expands. Comparisons of predicted jet $\Raa$ with experimental data from the LHC are also presented.
\end{abstract}

%
%
%
%
%
\section{Introduction}
In detail studies of the phenomenological aspects of jet quenching in RHIC and LHC have provided a better understanding of the dynamical processes involved in parton energy loss in recent years \cite{Baier:1994bd,Baier:1996sk,Zakharov:1996fv,Salgado:2003gb,Spousta:2015fca,Mehtar-Tani:2019ygg}. In fact, understanding the universal scaling features of radiative energy loss enables us to study the medium properties and scattering processes occurring at the partonic level. It has been well established that there is a scaling of the gluon spectra with average transport coefficient in an expanding medium as highlighted in \cite{Salgado:2003gb}. Furthermore, the single gluon brehmsstrahlung has been studied in detail for the case of time- dependent plasmas \cite{Arnold:2009mr}. On the other hand, medium modified gluon spectra have been solved analytically as well as numerically for the case of static plasmas \cite{Mehtar-Tani:2019ygg,Mehtar-Tani:2018zba}. Motivated with these results, we investigate the scaling properties for the re- summed gluon spectra in the expanding media through the kinetic rate equation.
 In this paper, we calculate the splitting rates for the formation of gluons inside expanding media from the single gluon emission spectrum. Next, we study the medium evolved spectra from the kinetic rate equation inside the medium  \cite{Mehtar-Tani:2014yea,Blaizot:2013vha,Blaizot:2013hx} . In order to study the multiple scattering processes within the media, we use the "harmonic oscillator" approximation  \cite{Baier:1994bd}.  We aim to quantify the scaling features with respect to different expansion profiles by evaluating the medium induced gluon spectra. Finally, we study the impact of the scaling features on the suppression factor of the jets for the case of expanding media. In this work, we report the differences in the scaling of the suppression from the average and optimal values of the quenching parameter for the expanding profiles. In a contemporary work \cite{Adhya:2019qse}, we have found that there is no universal way of scaling for the spectra over the full kinematical regime for the expanding profiles.

The paper is organized as follows.  In the next section, we highlight the important steps for the calculation of the gluon splitting rates from the single gluon emission spectra. Next, we numerically solve the kinetic rate equation using the gluon splitting rates derived from the spectra for different media as input kernels. Next, we present the quenching factor $Q_{AA}$ for different types of the media and discuss characteristic features seen in the spectra. Finally, we derive a scaling in the quenching parameter and  emphasize the impact of the medium evolution on the quenching factor. 
\section{Formalism}
First, let us briefly review the BDMPS-Z formalism for the medium induced gluon radiation from a static medium. This serves as a starting point for the discussion of the formalism which we will  extend to the case of medium induced gluon radiation for expanding media.

\subsection{In medium gluon spectra}
The transport coefficient for the static medium is  constant in time; $\hat q(t) = \hat q_0 $. 
In the multiple soft scattering approximation, the single gluon emission spectrum is given by \cite{Baier:1998yf,Salgado:2002cd, Arnold:2008iy},
\beq
\frac{\dd I}{\dd z}^{static} = \frac{\alpha_s}{\pi} P(z)\, \mathrm{Re} \ln \cos \Omega_0 L \,, \\
\eeq
where 
\beq
\label{eq:omega-static}
\Omega_0 L = \sqrt{\frac{-i}{2}\frac{\hat q_0}{p}}\kappa(z) L = \frac{1-i}{2} \kappa(z) \tau \,,
\eeq and $\kappa(z) = \sqrt{[1-z(1-z)]\big/[z (1-z)]}$. $L$ is the length of the medium at which the parton emerges into the vacuum.  In terms of the evolution variable $\tau \equiv L \sqrt{\hat{q_0}/p}$, the splitting rate is given by \cite{Adhya:2019qse},
\beq
\mathcal{K}(z,\tau)^{static}=\frac{dI}{dzd\tau} = \frac{\alpha_s}{2\pi} P(z) \kappa(z)\, \mathrm{Re} \left[(i-1) \tan \big((1-i)\kappa(z) \tau/2 \big) \right]\,.
\label{staticfullrate}
\eeq
We have only included the gluon splitting with the relevant Altarelli- Parisi splitting function.

Next, let us model the rapidly expanding medium as an exponentially evolving medium.
In this case, we can write the dynamically evolving transport coefficient as,
\beq
\hat q(t) = \hat q_0 {\rm e}^{-t/L}
\eeq
Using the above definition, the spectrum is given by \cite{Arnold:2008iy},
\beq
\frac{\dd I}{\dd z} ^{expo}= \frac{\alpha_s}{\pi} P(z)\, \mathrm{Re} \ln  J_0( 2\Omega_0 L ) \,, 
\eeq
where $\Omega_0$ is defined in Eq.[\ref{eq:omega-static}]. Using the single gluon spectra, we arrive at the expression for the splitting rate in the exponential medium as \cite{Adhya:2019qse},
\beq
\mathcal{K}(z,\tau)^{expo} = \frac{\alpha_s}{\pi} P(z) \kappa(z) \,\mathrm{Re} \left[ (i-1) \frac{J_1\big((1-i)\kappa(z) \tau \big)}{J_0\big((1-i)\kappa(z) \tau \big)} \right]
\label{exporate}
\eeq
Finally, we consider an uniformly expanding homogeneous quark gluon plasma formed in central nucleus nucleus collisions at time    $t=0$. At the formation time, characterised by $t_0$, the highly energetic parton radiates after leaving a hard collision occurring inside the medium at its highest density \cite{Arnold:2008iy,Salgado:2002cd}. Due to longitudinal expansion, the plasma cools down gradually following a Bjorken expansion power law \cite{Bjorken:1982qr}. This expansion enables to write the time dependent transport coefficient as,
\beq
\hat q(t) = \hat q_0 (t_0/t)^\alpha
\eeq
for $t_0 <t < t_0 +L$.
The power $\alpha$, for an ideal fluid, is chosen to be unity to model a simplistic hydro-dynamically expanding medium assuming no additional transverse expansion. On the other hand, setting $\alpha = 0$ gives us back the static case result characterized by a constant quenching.

Using the above expansion profile for the medium, the gluon emission spectrum  is given by \cite{Arnold:2008iy},
\beq
\frac{\dd I}{\dd z}^{BJ}&=& \frac{2\alpha_s}{\pi} P(z)\mathrm{Re} \ln\left[ \left(\frac{t_0}{L+t_0} \right)^{1/2} \frac{J_{\nu}(z_0)Y_{\nu-1}(z_L)-Y_{\nu}(z_0) J_{\nu-1}(z_L)}{J_\nu(z_L)Y_{\nu-1}(z_L) - Y_{\nu}(z_L) J_{\nu - 1}(z_L)} \right]\nonumber
\eeq
with $\nu \equiv 1/(2-\alpha)$ and
\begin{eqnarray}
z_0 &\equiv 2\nu \,\frac{1-i}{2} \kappa(z)\sqrt{\frac{\hat q_0}{p}} t_0 = \nu\, (1-i) \kappa(z) \tau_0 \,, \\
z_L &\equiv 2\nu\, \frac{1-i}{2} \kappa(z)\sqrt{\frac{\hat q_0}{p}} \sqrt{t_0 \,(L+t_0)} = \nu\, (1-i) \kappa(z) \sqrt{\tau_0 (\tau + \tau_0)} \,,
\end{eqnarray}
where $\tau_0 = \sqrt{\hat q_0/p} t_0$. For the concrete case of $\alpha = 1$ (and $\nu = 1$) \cite{Adhya:2019qse}, 
\beq
\frac{\dd I}{\dd z}^{BJ} = \frac{2\alpha_s}{\pi} P(z) \mathrm{Re} \ln\left[ \left(\frac{t_0}{L+t_0} \right)^{1/2} \frac{J_1(z_0)Y_0(z_L) - Y_1(z_0) J_0(z_L)}{J_1(z_L)Y_0(z_L) - Y_1(z_L) J_0(z_L)} \right]\, 
\eeq
Using the above equation, we derive the gluon splitting rate for the Bjorken expanding medium as,
\begin{eqnarray}
\mathcal{K}(z,\tau)^{Bjorken} &=& \frac{\alpha_s}{\pi} P(z) \kappa(z)\sqrt{\frac{\tau_0}{\tau + \tau_0}} \mathrm{Re} \nonumber\\
&&\times\left[ (1-i) \frac{J_1(z_L) Y_1(z_0) - J_1(z_0) Y_1(z_L) }{J_1(z_0) Y_0(z_L) - J_0(z_L) Y_1(z_0)} \right]
\label{bjorkenrate}
\end{eqnarray}

Since we have a constant (independent of time) quenching factor for the static medium, we obtain a constant rate for the same. The exponentially expanding medium has a rate that saturates to a constant at a later timescale. It should be noted that at large timescales, the factor containing the Bessel's arguments approaches unity, thereby saturating the rate to a constant value for the exponential profile. However, for the Bjorken expanding medium, although the factor containing the Bessel's arguments approaches unity at large timescales, the prefactor containing $\tau$ decreases the rate substantially at such timescales. The factor $P(z)\kappa(z)$ is a symmetric function in $z$ and remains common to all the profiles.

The parton cascade is implemented by the iteration of the single gluon emission spectra inside the kinetic rate equation which is given by,
  \begin{eqnarray}
  \label{eq:RateEquation-expanding-2}
  \frac{\partial D(x, \tau)}{\partial \tau} &= \int \dd z \,\mathcal{K}(z,\tau | p) \left[\sqrt{\frac{z}{x}} D\left(\frac{x}{z},\tau \right) - \frac{z}{\sqrt{x}} D(x,\tau) \right] \,,
  \end{eqnarray}
where 
\beq
\mathcal{K}(z,\tau|p) \equiv \sqrt{\frac{p}{\hat q_0}} \mathcal{K}(z,L|p) = \sqrt{\frac{p}{\hat q_0}} \frac{\dd I}{\dd z \dd L} \,.
\eeq
  The initial value of the $D(x,\tau)$ is a $\delta$-function in $x$ with a maximum for $x=1$ which characterizes the initial single color charge propagating trough the medium.
\begin{figure}[h]
\begin{center}
\begin{subfigure}[b]{1.0\textwidth}
\includegraphics[width=0.5\textwidth]{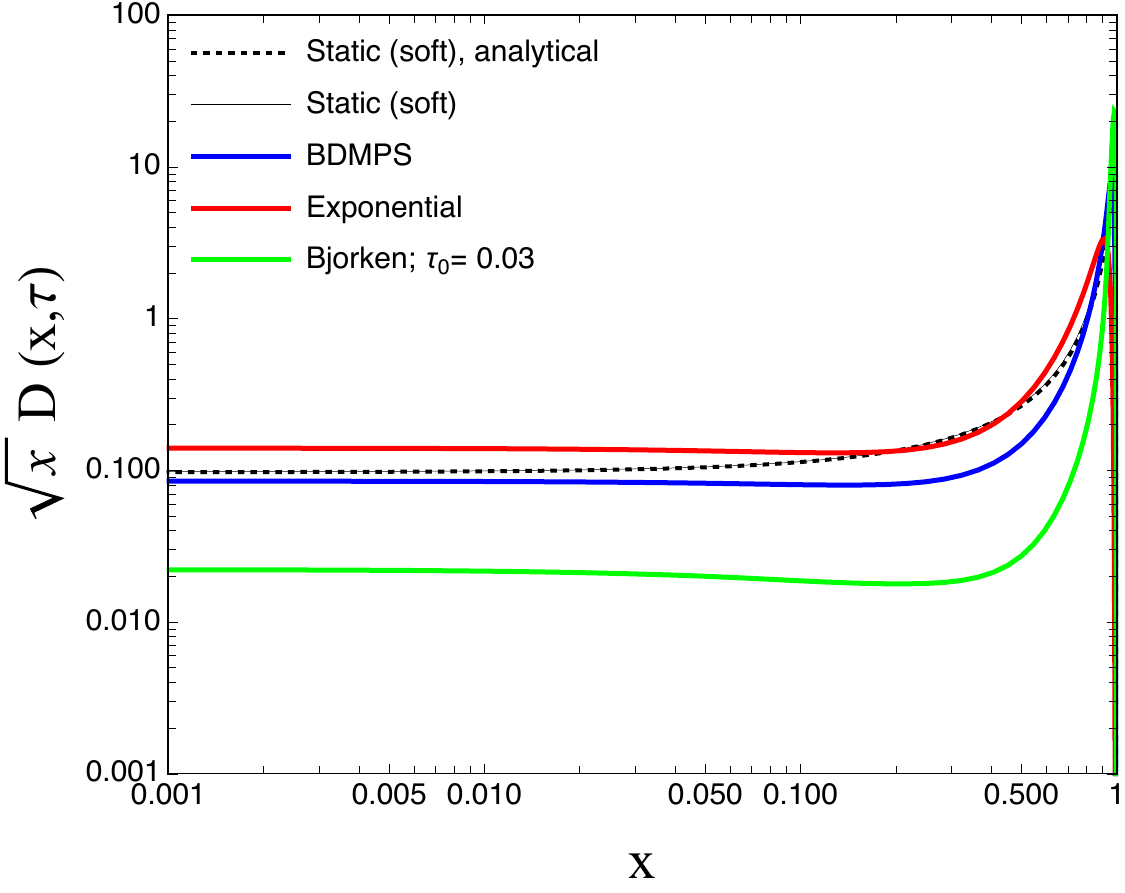}\includegraphics[width=0.5\textwidth]{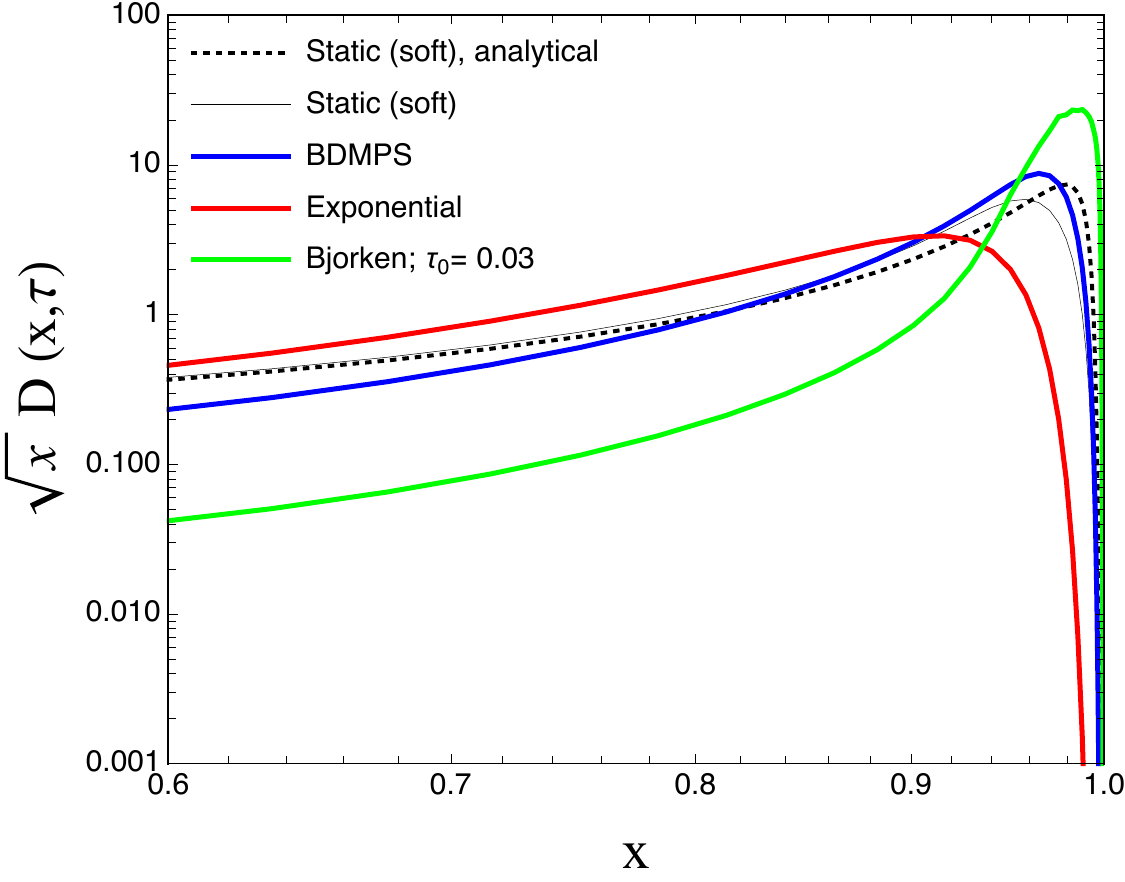}
\caption{ }
\end{subfigure}
\begin{subfigure}[b]{1.0\textwidth}
\includegraphics[width=0.5\textwidth]{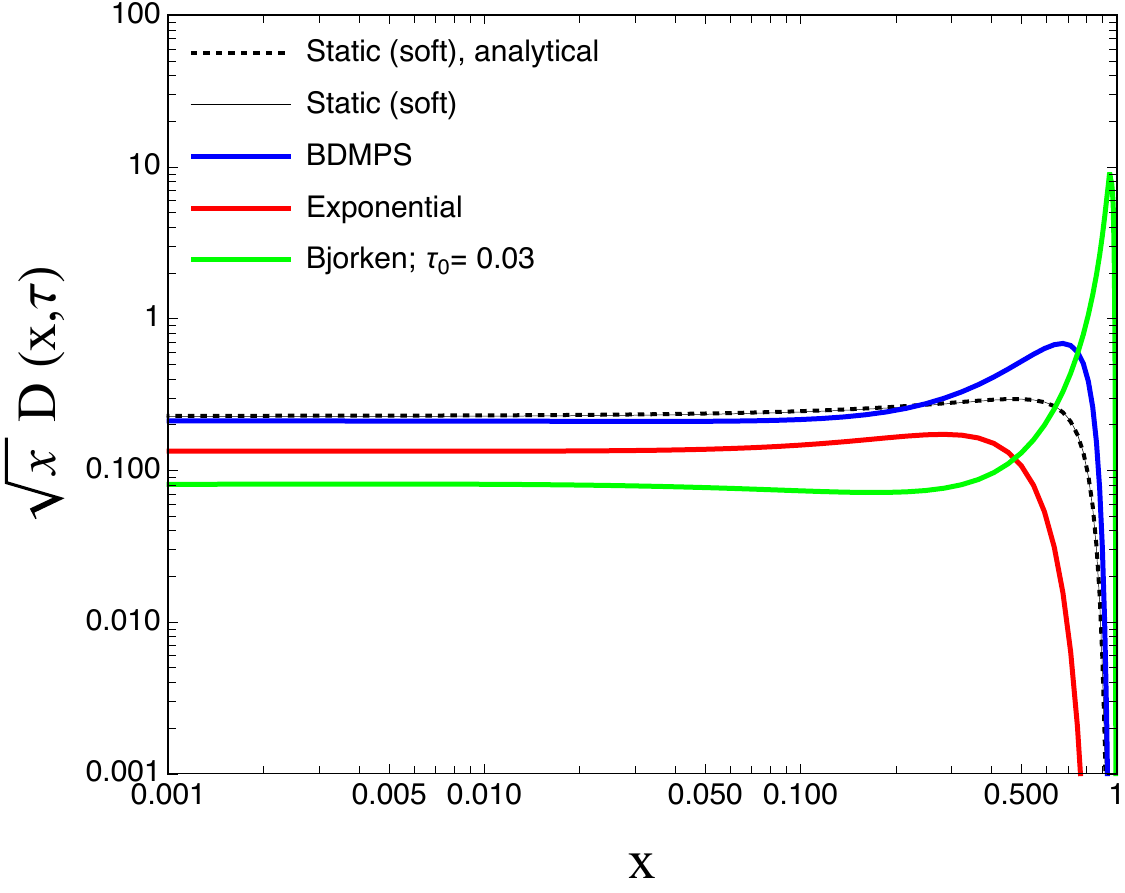}%
\includegraphics[width=0.5\textwidth]{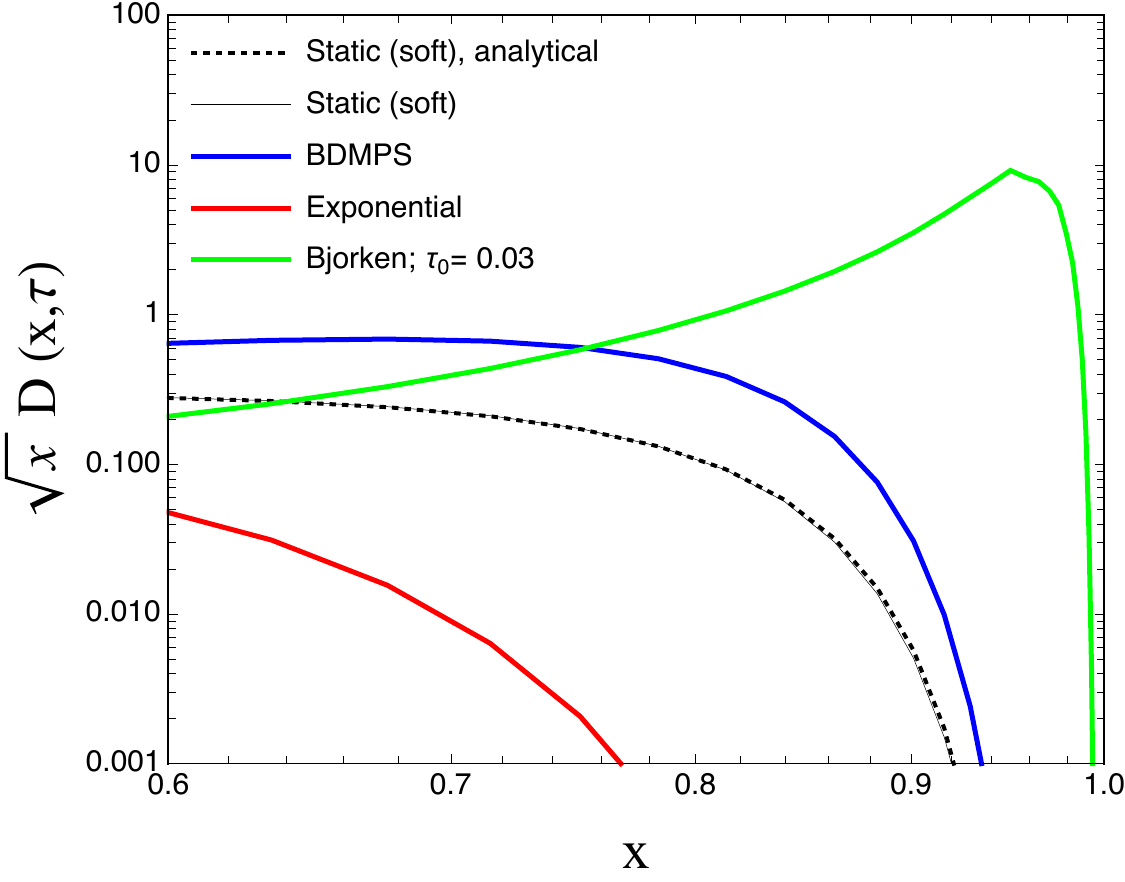}
\caption{ }
\end{subfigure}
\end{center}
\caption{The medium induced gluon spectra $D(x,\tau)$ as a function of the momentum fraction $x$ (with high $x$ region zoomed for figures on the right for both the panels) for $\tau = 0.1$ (panel (a)) and $\tau = 0.5$ (panel (b)) respectively. The static media has been plotted for soft approximation (black) and full kernel (blue) compared with the analytic soft static result (dashed black). The expanding cases have been plotted for exponential (red) and Bjorken (green) media respectively. }
\label{fig:spectrum-allx}
\end{figure}
In fig. \ref{fig:spectrum-allx}, we present the results for the medium evolved gluon spectra for the three kinds of media. In addition, we plot the analytic case of the "soft" part of the static spectra \cite{Mehtar-Tani:2013pia} and find that they agree with the numerical "soft" spectra. 
In this figure, for the evolution of the in- medium gluon cascade over $\tau$ (from 0.1 to 0.5), the $1/\sqrt{x}$ behavior of the spectrum occurs due to depletion of the energy flux at large $x \sim 1$ to small $x$ region of the spectra at different rates for different media. This depletes the original peak around $x \sim 1$ for different profiles at different timescales. In addition, the splitting rate for Bjorken profile is smallest for $x \rightarrow 1$ at any particular $\tau$ leading to slower quenching in comparison to all other profiles.
\subsection{Moments of the distribution and the nuclear modification factor}
\label{sec:moments}
\begin{figure}
\centering
\includegraphics[width=0.5\textwidth]{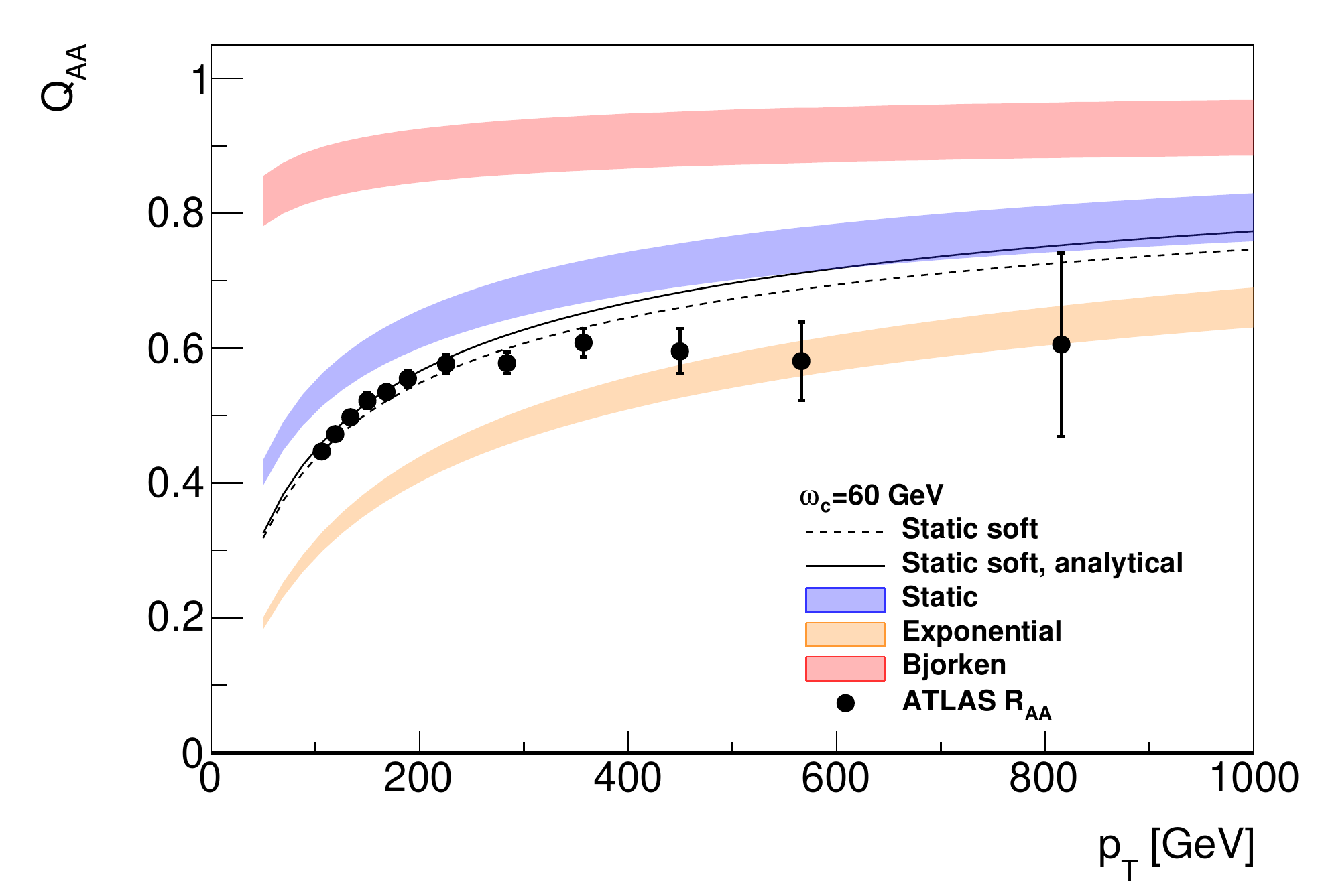}%
\includegraphics[width=0.5\textwidth]{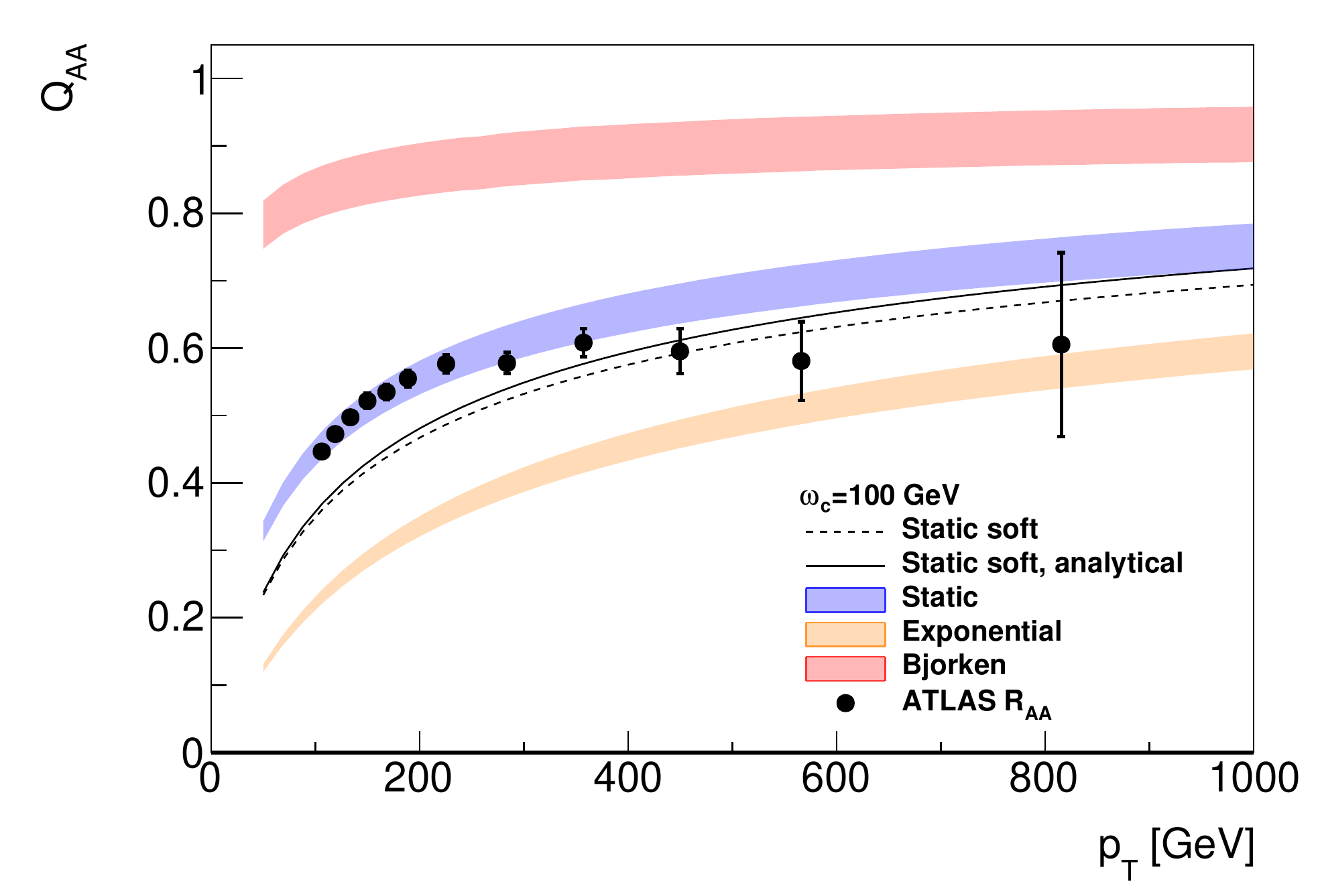}
\caption{The comparison of the jet suppression factor of the static media ( with full kernel (blue), soft kernel (black) and analytic soft result (dashed black)) with the expanding media (exponential (orange) and Bjorken (red)). The left and right figures corresponds to medium comparison at $\omega_c=60$~GeV and $\omega_c=100$~GeV respectively compared with the ATLAS data \cite{Aaboud:2018twu}.
 }
\label{fig:raa1}
\end{figure}
\begin{figure}
\centering
\includegraphics[width=0.49\textwidth]{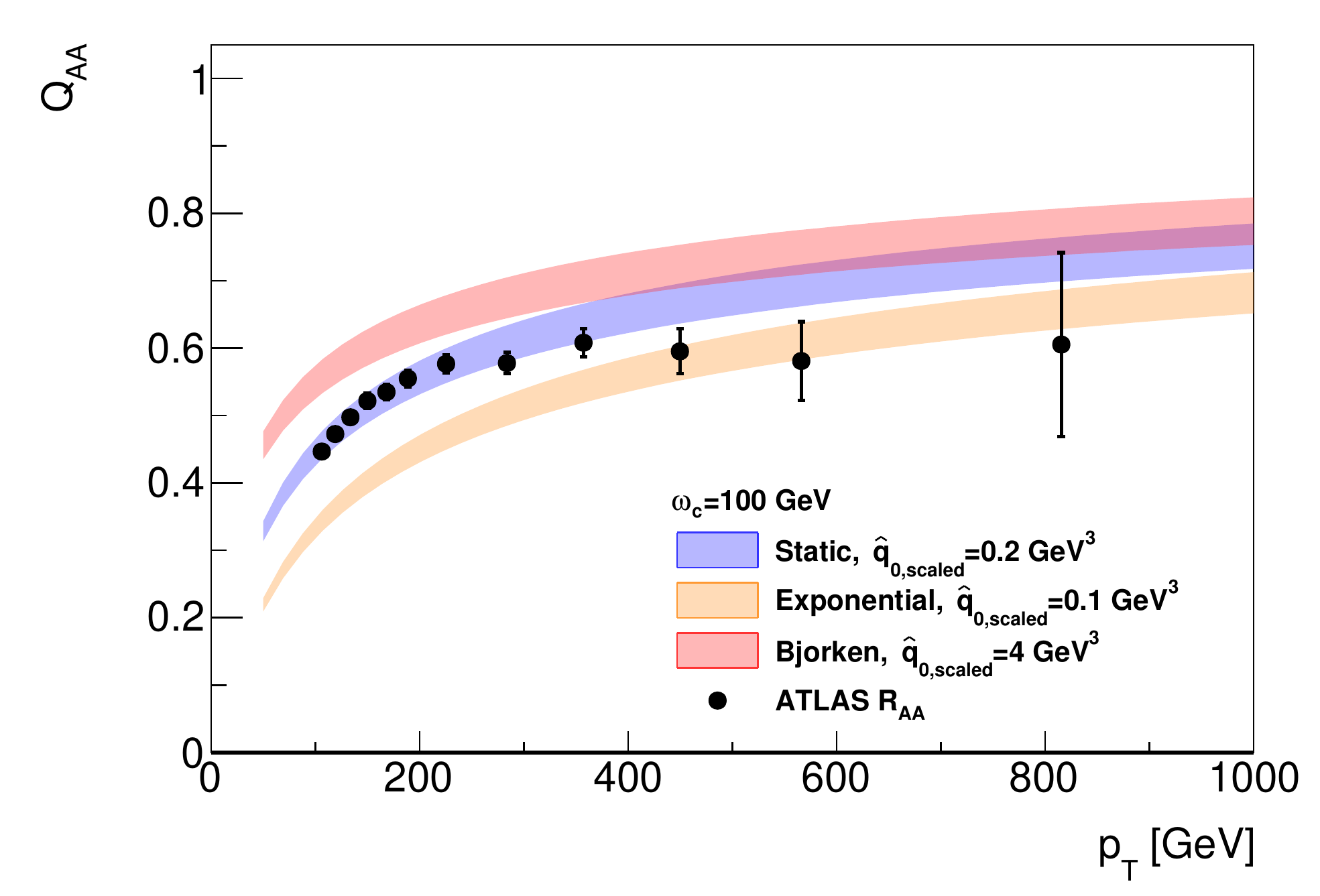}
\includegraphics[width=0.49\textwidth]{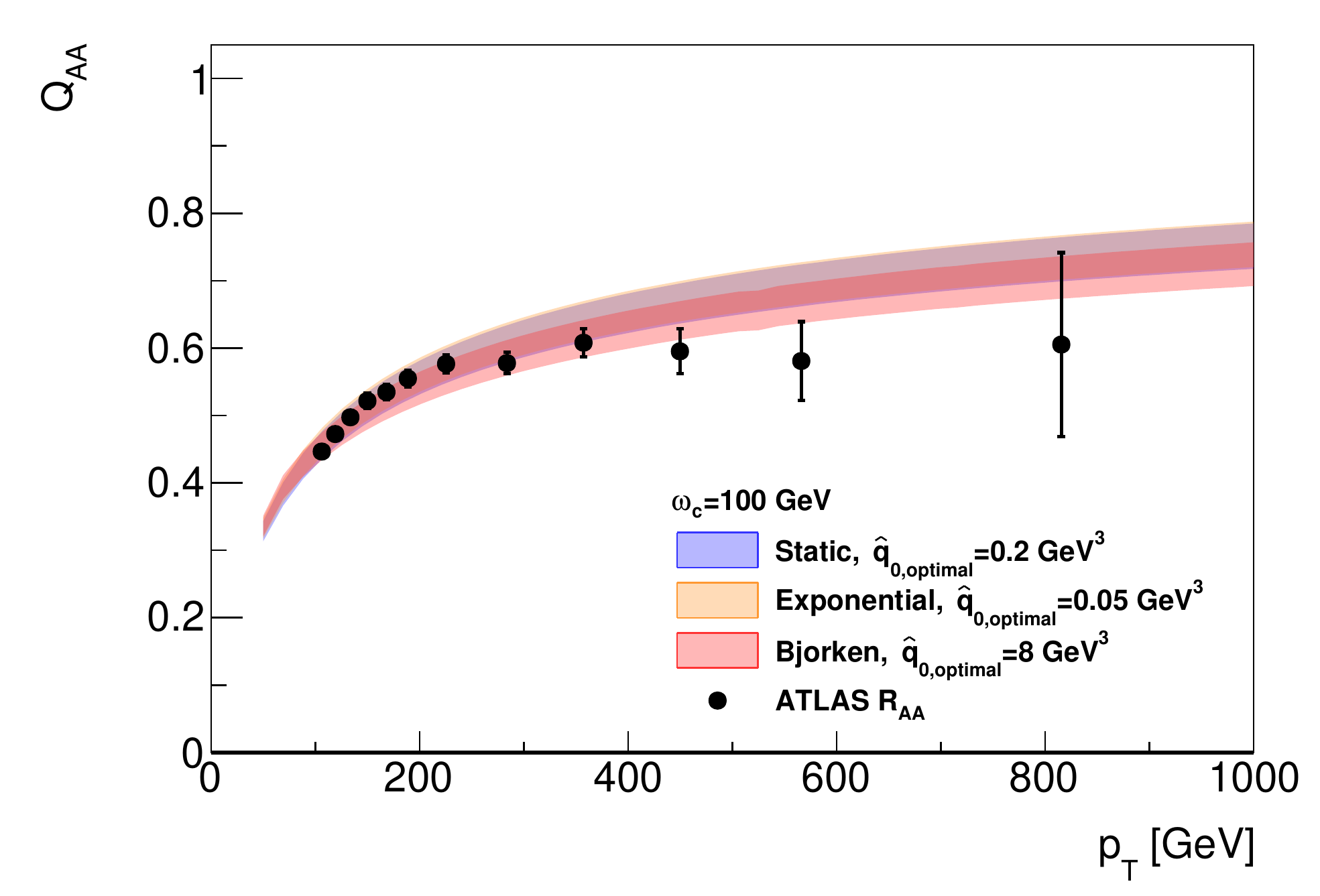}
\caption{A similar comparison of the different media profiles as that in fig.(\ref{fig:raa1}) at $\omega_c= 100$ $GeV$ with scaled average ($\hat{q}_{0,scaled}$) (left)  and optimal ($\hat{q}_{0,optimal}$) (right) scaling in the quenching parameter $\hat{q}$.
  }
\label{fig:raa2}
\end{figure}
The inclusive jet suppression in medium can be expressed in the form of convolution of the medium evolved gluon spectra $D(x,\tau)$ with the initial gluon spectra as,
  \beq
\frac{\dd \sigma_{AA}}{\dd \pT} = \int \dd \pT' \int_0^1 \frac{\dd x}{x} \,\delta(\pT- x \pT') D\left(x, \tau \equiv L \sqrt{\hat q / \pT'} \right) 
\frac{\dd \sigma_0}{\dd \pT'} \,.
\eeq
 The jet suppression factor $\Raa(\pT) = \dd \sigma_{\rm AA}/\dd \pT \Big/ \dd \sigma_0/\dd \pT$, is given by,
\beq
\Raa(\pT) = \int_0^1 \dd x \, x^{N-1} D(x, \sqrt{x} \tau) \,
\eeq
where we have assumed a power law approximation of the initial parton spectra given by $\dd \sigma_0/\dd \pT \propto \pT^{-N}$ with $N=5.6$. In our calculations, we limit to gluons as the only parton species. Therefore, $\Raa$ calculated here can be related to; although not equal to $R_{AA}$ as measured in experiments. In the case of the Bjorken model, the distribution has additionally a dependence on the initial evolution parameter $\tau_0$, $D(x,\tau_0,\tau)$, that is also rescaled by the (unknown) initial $\pT$, hence we obtain
\beq
\Raa(\pT) = \int_0^1 \dd x \, x^{N-1} D(x,\sqrt{x} \tau_0, \sqrt{x} \tau) \,.
\eeq

Fig.\ref{fig:raa1} shows the resulting $Q_{AA}$ distributions for different types of the expansion for two choices of  $\omega_c$; $60$~GeV and $100$~GeV respectively. Here, $\omega_c (=\hat{q}L^2/2)$ is the characteristic gluon frequency for a hard parton traversing the finite medium of length $L$. 
 We observe a large difference for different kinds of media. It was shown in Ref.~\cite{Salgado:2003gb}, that the impact of the difference in the type of the medium expansion on the $\dd I/ \dd z$ can be 
scaled out by replacing the $\hat{q}$ parameter by the average, $\qavg$. Left panel of Fig.\ref{fig:raa2} shows this configuration ($\omega_c \rightarrow \bar{\omega_c} 
= \omega_c/\qavg$) with scaling of the expanding quenching factors to the static one defined by $\hat{q}_{0,scaled}$ (average scaling) and determined to be approximately $0.2$, $0.1$, and $4$ $GeV^3$ for the static, exponential, and Bjorken type of the expansion, 
respectively. We observe that the difference among the different medium profiles arises due to the difference of the dependencies of the single gluon emission emission spectra $\dd I/ \dd z$ on $\tau$. However, medium evolution of the spectra has considerable effects as a consequence of which the average quenching parameter $\qavg$ cannot account for 
the different types of the expansion. Finally, we are able to derive an optimal quenching parameter, $\hat{q}_{0,optimal}$, which minimizes the difference between the 
data and theoretical estimation. The result of this minimization is shown in the right panel of fig.(\ref{fig:raa2}) with $\hat{q}_{0,optimal}$ being approximately $0.2$, $0.05$, and $8$ $GeV^3$ for 
static, exponential and Bjorken type of the expansion respectively. This large difference among the quenching parameter implies that the appropriate modeling of the medium 
expansion is an important component of the effort to understand the jet quenching mechanism.
\section{Conclusions}
\label{sec:conclusions}
In this work, we have studied the effect of the expanding QGP media on the gluon spectra in multiple soft approximation. We have considered models of the static, exponential and Bjorken expanding media. Rates of gluon emission were evaluated from single gluon spectra and medium evolved spectra were calculated by means of numerical solutions of 
the evolution equation for the gluon radiation. Further, medium evolved gluon spectra are  used to calculate the quenching factor of jets. We find that the 
impact of the medium expansion cannot be scaled out only by the average quenching parameter of $\hat q$. Values of quenching parameter which minimize the difference between the 
data and theoretical estimates are derived for different types of medium expansion.
  Large difference between these values implies that use of realistic parameterization (such as inclusion of flavor dependent cascades) of the medium expansion is needed to improve on the precision of modeling of the 
jet quenching phenomenon.

\section*{Acknowledgments}

  KT is supported by a Starting Grant from Trond Mohn Foundation (BFS2018REK01) and the University of Bergen. 
  CAS is supported by Ministerio de Ciencia e Innovaci\'on of Spain under project FPA2017-83814-P; Unidad de Excelencia Mar\'ia de Maetzu under project MDM-2016-0692;
ERC-2018-ADG-835105 YoctoLHC; and Xunta de Galicia (Conseller\'ia de Educaci\'on) and FEDER.
  SPA and MS are supported by Grant Agency of the Czech Republic under grant 18-12859Y, by the Ministry of Education, Youth and Sports of the Czech Republic under grant 
LTT~17018, and by Charles University grant UNCE/SCI/013.

\section*{References}

\bibliographystyle{unsrt}
\bibliography{icnfp_proc2019}

\begin{thebibliography}{10}

\bibitem{Baier:1994bd}
R.~Baier, Yuri~L. Dokshitzer, S.~Peigne, and D.~Schiff.
\newblock {Induced gluon radiation in a QCD medium}.
\newblock {\em Phys.Lett.}, B345:277--286, 1995.

\bibitem{Baier:1996sk}
R.~Baier, Yuri~L. Dokshitzer, Alfred~H. Mueller, S.~Peigne, and D.~Schiff.
\newblock {Radiative energy loss and p(T) broadening of high-energy partons in
  nuclei}.
\newblock {\em Nucl.Phys.}, B484:265--282, 1997.

\bibitem{Zakharov:1996fv}
B.G. Zakharov.
\newblock {Fully quantum treatment of the Landau-Pomeranchuk-Migdal effect in
  QED and QCD}.
\newblock {\em JETP Lett.}, 63:952--957, 1996.

\bibitem{Salgado:2003gb}
Carlos~A. Salgado and Urs~Achim Wiedemann.
\newblock {Calculating quenching weights}.
\newblock {\em Phys.Rev.}, D68:014008, 2003.

\bibitem{Spousta:2015fca}
Martin Spousta and Brian Cole.
\newblock {Interpreting single jet measurements in Pb $+$ Pb collisions at the
  LHC}.
\newblock {\em Eur. Phys. J.}, C76(2):50, 2016.

\bibitem{Mehtar-Tani:2019ygg}
Yacine Mehtar-Tani and Konrad Tywoniuk.
\newblock {Improved opacity expansion for medium-induced parton splitting}.
\newblock {\em JHEP}, 06:187, 2020.

\bibitem{Arnold:2009mr}
Peter~Brockway Arnold.
\newblock {High-energy gluon bremsstrahlung in a finite medium: harmonic
  oscillator versus single scattering approximation}.
\newblock {\em Phys. Rev.}, D80:025004, 2009.

\bibitem{Mehtar-Tani:2018zba}
Yacine Mehtar-Tani and Soeren Schlichting.
\newblock {Universal quark to gluon ratio in medium-induced parton cascade}.
\newblock {\em JHEP}, 09:144, 2018.

\bibitem{Mehtar-Tani:2014yea}
Yacine Mehtar-Tani and Konrad Tywoniuk.
\newblock {Jet (de)coherence in PbPb collisions at the LHC}.
\newblock {\em Phys. Lett.}, B744:284--287, 2015.

\bibitem{Blaizot:2013vha}
Jean-Paul Blaizot, Fabio Dominguez, Edmond Iancu, and Yacine Mehtar-Tani.
\newblock {Probabilistic picture for medium-induced jet evolution}.
\newblock {\em JHEP}, 06:075, 2014.

\bibitem{Blaizot:2013hx}
Jean-Paul Blaizot, Edmond Iancu, and Yacine Mehtar-Tani.
\newblock {Medium-induced QCD cascade: democratic branching and wave
  turbulence}.
\newblock {\em Phys. Rev. Lett.}, 111:052001, 2013.

\bibitem{Adhya:2019qse}
Souvik~Priyam Adhya, Carlos~A. Salgado, Martin Spousta, and Konrad Tywoniuk.
\newblock {Medium-induced cascade in expanding media}.
\newblock {\em JHEP}, 07:150, 2020.

\bibitem{Baier:1998yf}
R.~Baier, Yuri~L. Dokshitzer, Alfred~H. Mueller, and D.~Schiff.
\newblock {Radiative energy loss of high-energy partons traversing an expanding
  QCD plasma}.
\newblock {\em Phys. Rev.}, C58:1706--1713, 1998.

\bibitem{Salgado:2002cd}
Carlos~A. Salgado and Urs~Achim Wiedemann.
\newblock {A Dynamical scaling law for jet tomography}.
\newblock {\em Phys.Rev.Lett.}, 89:092303, 2002.

\bibitem{Arnold:2008iy}
Peter~Brockway Arnold.
\newblock {Simple Formula for High-Energy Gluon Bremsstrahlung in a Finite,
  Expanding Medium}.
\newblock {\em Phys.Rev.}, D79:065025, 2009.

\bibitem{Bjorken:1982qr}
J.D. Bjorken.
\newblock {Highly Relativistic Nucleus-Nucleus Collisions: The Central Rapidity
  Region}.
\newblock {\em Phys.Rev.}, D27:140--151, 1983.

\bibitem{Mehtar-Tani:2013pia}
Yacine Mehtar-Tani, Jose~Guilherme Milhano, and Konrad Tywoniuk.
\newblock {Jet physics in heavy-ion collisions}.
\newblock {\em Int. J. Mod. Phys.}, A28:1340013, 2013.

\bibitem{Aaboud:2018twu}
{ATLAS Collaboration}.
\newblock {Measurement of the nuclear modification factor for inclusive jets in
  Pb+Pb collisions at $\sqrt{s_\mathrm{NN}}=5.02$ TeV with the ATLAS detector}.
\newblock {\em Phys. Lett.}, B790:108--128, 2019.

\end{thebibliography}

\end{document}